\documentclass{article}

\usepackage{PRIMEarxiv}

\usepackage[utf8]{inputenc} 
\usepackage[T1]{fontenc}    
\usepackage{hyperref}       
\usepackage{url}            
\usepackage{booktabs}       
\usepackage{amsfonts}       
\usepackage{nicefrac}       
\usepackage{microtype}      
\usepackage{amsmath}
\usepackage{algorithmic}
\usepackage{array}
\usepackage{lipsum}
\usepackage{multirow}
\usepackage{tabularx}
\usepackage{fancyhdr}       
\usepackage{graphicx}       
\graphicspath{{media/}}     
\usepackage{color}

\title{Head-Neck Dual-energy CT Contrast Media Reduction Using Diffusion Models
}

\author{
  Qing Lyu$^*$, Josh Tan, Megan E. Lipford, Micheal E. Zapadka, Christopher M. Lack, Christopher T. Whitlow$^*$ \\
  Wake Forest University School of Medicine \\
  Winston-Salem, NC \\
  \texttt{\{qlyu, jtan, m.lipford, mzapadka, clack, cwhitlow\}@wakehealth.edu}
  \And
  Jonathan D. Clemente, \\
  Carolinas Medical Center \\
  Charlotte, NC \\
  \texttt{Jonathan.Clemente@atriumhealth.org}
  \And
  Chuang Niu, Ge Wang$^*$ \\
  Rensselaer Polytechnic Institute \\
  Troy, NY\\
  \texttt{\{niuc, wangg6\}@rpi.edu} \\
\\
$^*$Corresponding authors
}

\begin{document}
\maketitle

\begin{abstract}
Iodinated contrast media is essential for dual-energy computed tomography (DECT) angiography. Previous studies show that iodinated contrast media may cause side effects, and the interruption of the supply chain in 2022 led to a severe contrast media shortage in the US. Both factors justify the necessity of contrast media reduction in relevant clinical applications. In this study, we propose a diffusion model-based deep learning framework to address this challenge. First, we simulate different levels of low contrast dosage DECT scans from the standard normal contrast dosage DECT scans using material decomposition. Conditional denoising diffusion probabilistic models are then trained to enhance the contrast media and create contrast-enhanced images. Our results demonstrate that the proposed methods can generate high-quality contrast-enhanced results even for images obtained with as low as 12.5\% of the normal contrast dosage. Furthermore, our method outperforms selected competing methods in a human reader study. 
\end{abstract}


\section{Introduction}
Dual-energy computed tomography (DECT) is an advanced imaging technique that captures projection data with two energy spectra simultaneously or almost simultaneously. Because the x-ray linear attenuation coefficients of tissues and materials varies with the X-ray energy level, DECT can differentiate material components and quantify them tomographically~\cite{de2012dual,liu2009quantitative}. In the last decade, DECT has been widely used in many applications especially oncology to enhance tumor detection and characterization~\cite{roele2017dual,forghani2015advanced}.

Intravenous iodinated contrast media is essential for DECT angiography to  analyze pathology and monitor treatment. While it is generally considered safe, previous studies show that its use may result in anaphylactoid and/or non-anaphylactoid adverse reactions, some of which may even be life-threatening~\cite{singh2008iodinated}. Acute kidney injury is a common iodinated contrast-induced problem~\cite{mccullough2008acute}, and there are growing concerns about its use. On the one hand, the number of CT scans performed annually in the US has increased significantly over the past few decades, with a large percentage of these scans using intravenous contrast media~\cite{brenner2010slowing}. On the other hand, as the patient population ages, chronic kidney disease and diabetes become more common, and patients are exposed to more iodinated contrast media, contrast-induced acute kidney injury is likely to become an even more significant challenge in the future~\cite{mccullough2008acute,haubold2021contrast}.

Last year, the COVID-19-related shutdown of a General Electric Healthcare factory in Shanghai disrupted the production of iodinated contrast media, leading to a major shortage in the US~\cite{contrastmedialshortage,davenport2022comparison}. Although the shortage has been over now, there is a possibility that it could recur due to the uncertainties related to future pandemics, supply chains, other factors. This further exacerbates the concern over the routine use of iodinated contrast media. 

Reducing the clinical usage of iodinated contrast media in DECT scans offers several advantages. First, limiting its use would be more patient-friendly and reduce the risk of contrast media-induced adverse reactions, particularly for patients in critical condition or those who need repeated contrast administration~\cite{mccullough2008acute}. Second, it significantly reduces the costs associated with contrast media usage~\cite{robinson2013evaluating,sharma2008iodinated}. According to~\cite{contrastmedialmarket}, the global market for contrast media is estimated to reach \$5.4 billion by 2026, indicating a significant financial impact of its usage. Third, it is a viable solution to address the potential contrast media shortage in the future. 

In recent years, efforts were made to reduce the needed dose of iodinated contrast media by lowering the X-ray tube voltage. It was reported that scanning patients at low tube voltage settings (such as 80 kVp) with a reduced iodinated contrast media dose (around 50\% to 60\% of the normal dose) can generate images without significant quality degradation after iterative reconstruction~\cite{buls2015contrast,nakaura2011abdominal,sigal2004low,hunsaker2010contrast,higashigaito2016ct}. Recently, deep learning has produced promising results in several medical imaging applications such as CT denoising and metal artifact reduction~\cite{yang2018low,niu2021multiple}. In the MRI domain, due to safety concerns about gadolinium-based contrast media, multiple deep learning methods were proposed to generate synthetic images that are comparable to full dose images using 10\% or even zero contrast media~\cite{gong2018deep,kleesiek2019can,luo2021deep,bone2021contrast,montalt2021reducing,pasumarthi2021generic}. For CT iodinated contrast media reduction, Haubold \textit{et al.} proposed a modified Pix2PixHD model that can generate contrast-enhanced images of comparable image quality and diagnostic accuracy from images with 50\% contrast media reduction~\cite{haubold2021contrast}. Liu \textit{et al.} developed a DyeFreeNet
model that can generate virtual contrast-enhanced CT images from non-contrast CT scans~\cite{liu2020dyefreenet}.

In this study, we propose an iodinated contrast media reduction method based on denoising diffusion probabilistic models (DDPMs)~\cite{ho2020denoising}. Compared to other contrast-enhancement methods such as material decomposition enhancement (MDE) and UNet, our proposed method can generate superior results with better representation of contrast-enhanced vessels in head and neck DECT images using as low as 12.5\% of the normal iodinated contrast dose. 

\begin{figure*}[!ht]
\centering
\includegraphics[width=\textwidth]{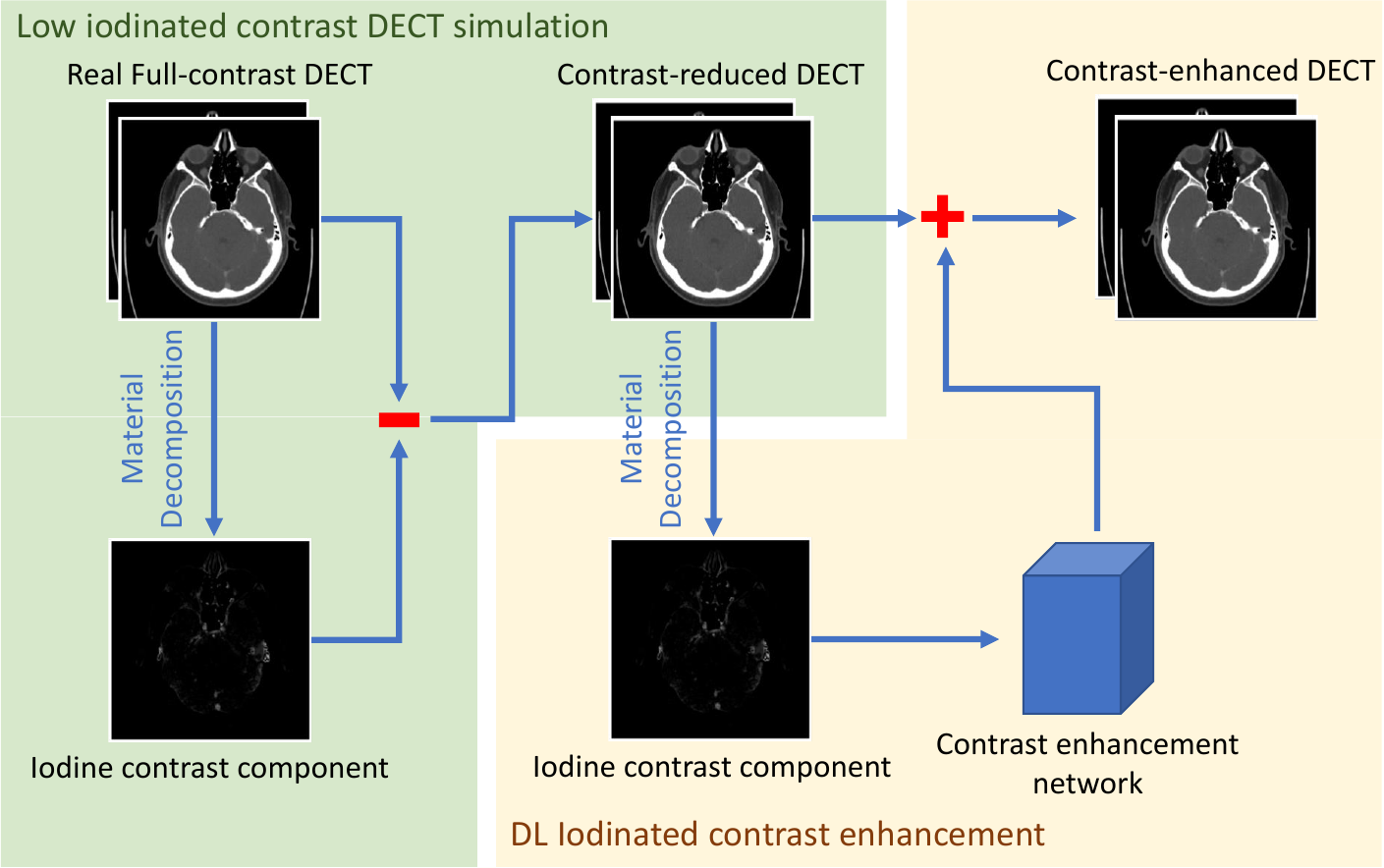}
\caption{Overview of our proposed DDPM-based contrast enhancement method.}
\label{fig_1}
\end{figure*}

\section{Methodology}
As shown in Fig.~\ref{fig_1}, the proposed method consists of the two components: the low iodinated contrast DECT simulation and the deep learning iodinated contrast enhancement. The low iodinated contrast DECT simulation realistically simulates low-contrast DECT scans as training data, while the deep learning iodinated contrast enhancement users a trained deep network to generate contrast-enhanced results from a low iodinated contrast dose DECT scan.

\subsection{Data Acquisition}
We used a total of 110 DECT scans collected by the Atrium Health Wake Forest Baptist Medical Center in May and June, 2022 on Siemens Somatom Drive CT scanners under the head or head \& neck DECT angiography protocols. In the imaging process, Omnipaque iohexol contrast media was injected, with the amount of iodinated contrast media used for each patient ranging between 50 $cc$ and 80 $cc$, as indicated by the medications list in the medical record. The Syngo.via server was used to produce monoenergetic images from the original DECT scans in the Monoenergetic plus setting. After processing, we obtained 80 $keV$ and 110 $keV$ monoenergetic images from each DECT scan, with a slice thickness of 0.6 $mm$ and pixel resolution of 0.46 $mm$ by 0.46 $mm$. 80 scans were used for the training and 20 scans were used for testing. Another 10 scans were used for reader study. 

\subsection{Material decomposition}
As in our previous paper~\cite{cong2020virtual}, we used the three-material mass fraction decomposition algorithm~\cite{liu2009quantitative} based on 80 $keV$ and 110 $keV$ monoenergetic images. Through material decomposition, each monoenergetic image was decomposed into the summation of three basis material maps: water, bone, and iodinated contrast media. We calculated the linear attention coefficients of water, bone, and iohexol contrast media based on the mass attenuation coefficients and mass densities reported by National Institute of Standards and Technology~\cite{hubbell1995tables} and {\c{C}}ak{\i}r~\textit{et al.}~\cite{ccakir2020determining}. The mass fraction of each material can be calculated using the following equations:
\begin{equation}\label{equ_1}
\Biggl\{ \begin{array}{lcl} \mu(E_{L}) = v_{a}\mu_{a}(E_{L}) + v_{b}\mu_{b}(E_{L}) + v_{c}\mu_{c}(E_{L}) \\ \mu(E_{H}) = v_{a}\mu_{a}(E_{H}) + v_{b}\mu_{b}(E_{H}) + v_{c}\mu_{c}(E_{H}) \\ 1 = v_{a} + v_{b} + v_{c} \end{array}, 
\end{equation}
where $\mu(E_{L})$ and $\mu(E_{H})$ represent the averaged pixel linear attenuation coefficients in the 80 $keV$ low-energy and 110 $keV$ high-energy images respectively. The linear attenuation coefficients of water, bone, and  contrast media under the 80 $keV$ low-energy and 110 $keV$ high-energy settings, denoted as $\mu_{a}(E_{L}), \mu_{b}(E_{L}), \mu_{c}(E_{L}), \mu_{a}(E_{H}), \mu_{b}(E_{H})$, and $\mu_{c}(E_{H})$, are listed in Table~\ref{tab_1}. $v_{a}, v_{b}$, and $v_{c}$ represent the mass fractions of the three materials.

\begin{table}
 \caption{Linear attenuation coefficients of basis materials.}
  \centering
  \begin{tabular}{c c c c}
    \toprule
    \multirow{2}{*}{Energy ($keV$)} & \multicolumn{3}{c}{Attenuation ($cm^{-1}$)} \\
    \cmidrule(r){2-4} 
     &  Water & Bone & Iohexol \\
    \midrule
    80 & 0.184 & 0.428 & 3.784 \\
    110 & 0.167 & 0.342 & 2.066 \\
    \bottomrule
  \end{tabular}
  \label{tab_1}
\end{table}

\subsection{Synthesizing multi-level contrast dose reduced images}
After we obtained the contrast media mass fraction map, we generated synthesized multi-level contrast media monoenergetic images by either weakening or strengthening the contrast media component. For each 80 $keV$ or 110 $keV$ monoenergetic scan, we generated synthesized images with contrast dose levels of 10 $cc$, 20 $cc$, 30 $cc$, 40 $cc$, 50 $cc$, 60 $cc$, 70 $cc$, and 80 $cc$. In this study, we considered 80 $cc$ images as the full contrast dose ground truths. All other contrast dose level images were processed using the proposed method to produce contrast media enhanced images. It is worth noting that for the 10 $cc$, 20 $cc$, 30 $cc$, and 40 $cc$ contrast dose level image enhancement, all images used were synthesized images. Conversely, for the 50 $cc$, 60 $cc$, and 70 $cc$ contrast dose level image enhancement, we used a combination of synthesized images and real low contrast dose level images. 

\subsection{Contrast enhancement network}
Here we adapted the conditional denoising diffusion probabilistic model (DDPM) for contrast media enhancement. DDPM, a recently established generative model~\cite{ho2020denoising}, has shown great successes in generating high-quality natural or artistic images~\cite{rombach2022high}. We optimized a customized DDPM for our medical imaging task: employing a low iodinated contrast DECT scan as the input to generate a contrast-enhanced counterpart. The backbone of the DDPM is a UNet~\cite{ronneberger2015u} architecture, which is consistent with the original DDPM literature.

\begin{figure*}[!ht]
\centering
\includegraphics[width=\textwidth]{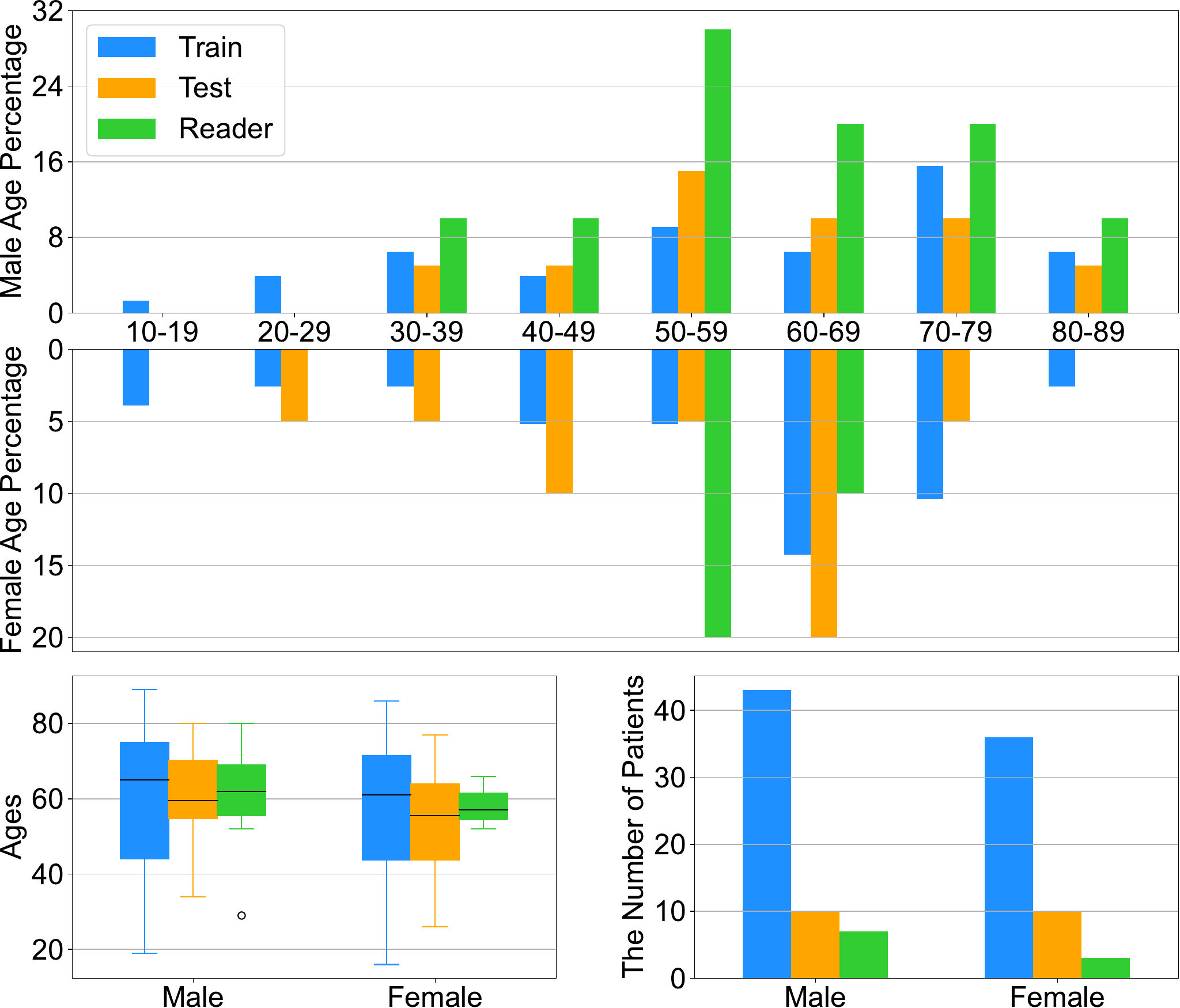}
\caption{Patient demographic comparison of DECT scans in the training, testing, and reader study datasets.}
\label{fig_2}
\end{figure*}

\subsection{Reader Study}
To evaluate the quality of different contrast enhancement results, we invited three radiologists (28, 21, and 15 years of experience) to review each result and give a score on the five-point scale with 1-point for the worst and 5-point for the best. The reader study was conducted under two settings: a extremely low contrast dosage (10 $cc$) setting and a half contrast dosage (40 $cc$) setting. Under each setting, the scores were obtained on the quality of normal contrast dosage images, simulated low contrast dosage images, and contrast enhanced results obtained using three contrast enhancement methods (MDE, UNet, and DDPM) respectively. For the 10 cases used for this reader study, there are 5 scans without abnormality and 5 scans with abnormity.

\section{Results}
\subsection{Data demographic analysis}
In this study, 80 DECT scans were used for training, another 20 scans for testing, and the remaining 10 scans for the reader study. Fig.~\ref{fig_2} shows the statistics of patient scans used for training, testing, and reader study respectively. In the training dataset, there are 37 female patients and 43 male patients with an overall average age of 58.4 years old. The testing dataset is made up of 10 male patients and 10 female patients with an average age of 56.6 years old. The reader study dataset consists of 7 male patients and 3 female patients with an average age of 59.5 years old.

\subsection{DDPM Contrast Enhancement}
\begin{figure*}[!ht]
\centering
\includegraphics[width=\textwidth]{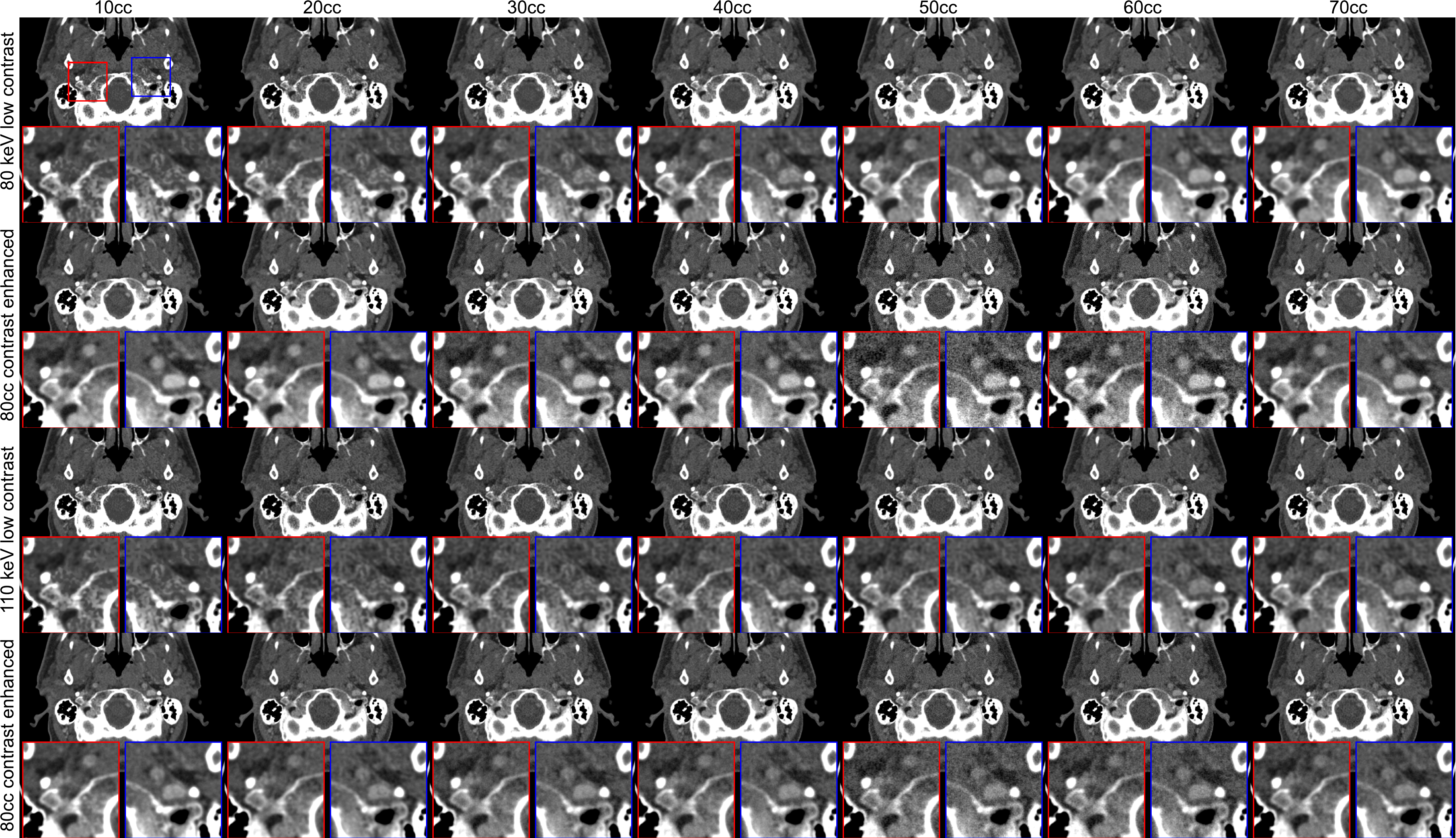}
\caption{Comparison between the contrast-enhanced results and their corresponding simulated low contrast dosage images. The top half images displays the results obtained at 80 $keV$, while the bottom shows the results obtained at 110 $keV$.}
\label{fig_3}
\end{figure*}

We compared simulated low contrast dosage images and their corresponding contrast enhanced images in Fig.~\ref{fig_3}. It can be seen that for both 80 $keV$ and 110 $keV$ images, the visibility of vessels gradually diminished as the dosage of contrast media was decreased. In the 10 $cc$ and 20 $cc$ images, it became difficult to detect the vessels due to their vague appearance. In contrast, our contrast enhancement results effectively highlighted the vessels under all conditions. Taking the 10 $cc$ enhancement as an example, our method successfully defined the contrast-enhanced vessels, whereas the original low dosage images failed to display them distinctly.

\subsection{Comparison of Contrast Enhancements Using Different Methods}
\begin{figure*}[!ht]
\centering
\includegraphics[width=\textwidth]{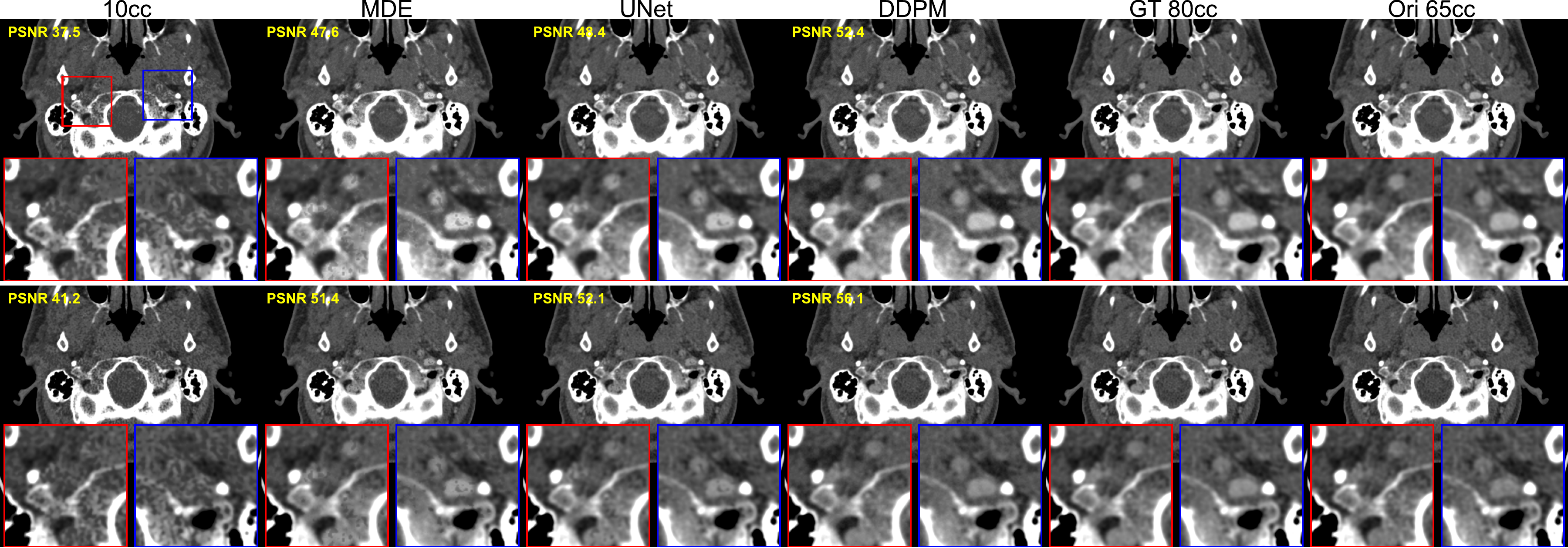}
\caption{Comparison between MDE, UNet, our contrast-enhancement results. The first column shows 10 $cc$ low contrast dose DECT images. The second, third, and fourth columns contain MDE, UNet, and our contrast-enhancement results. The fifth and sixth columns present 80 $cc$ normal contrast dosage ground truth and the original 65 $cc$ DECT scan images, respectively.}
\label{fig_4}
\end{figure*}

We further compared our results with those obtained using other methods such as MDE and UNet. The MDE method is a non-deep learning approach that directly focuses on enhancing the contrast component obtained from the material decomposition of low contrast dose DECT images. On the other hand, UNet is a widely used deep learning model successful in various medical image analysis tasks. As illustrated in Fig.~\ref{fig_4}, all the three methods are capable of presenting contrast-enhanced vessels with clearer shapes and better-defined boundaries. However, when comparing our results and the ground truth to the MDE and UNet results, it can be observed that the enhanced vessels in MDE and UNet images had some small dark holes. In contrast, our results did not suffer from these holes, delineating vessels with uniform pixel intensity and closely resembling the ground truth.

\begin{figure*}[!ht]
\centering
\includegraphics[width=\textwidth]{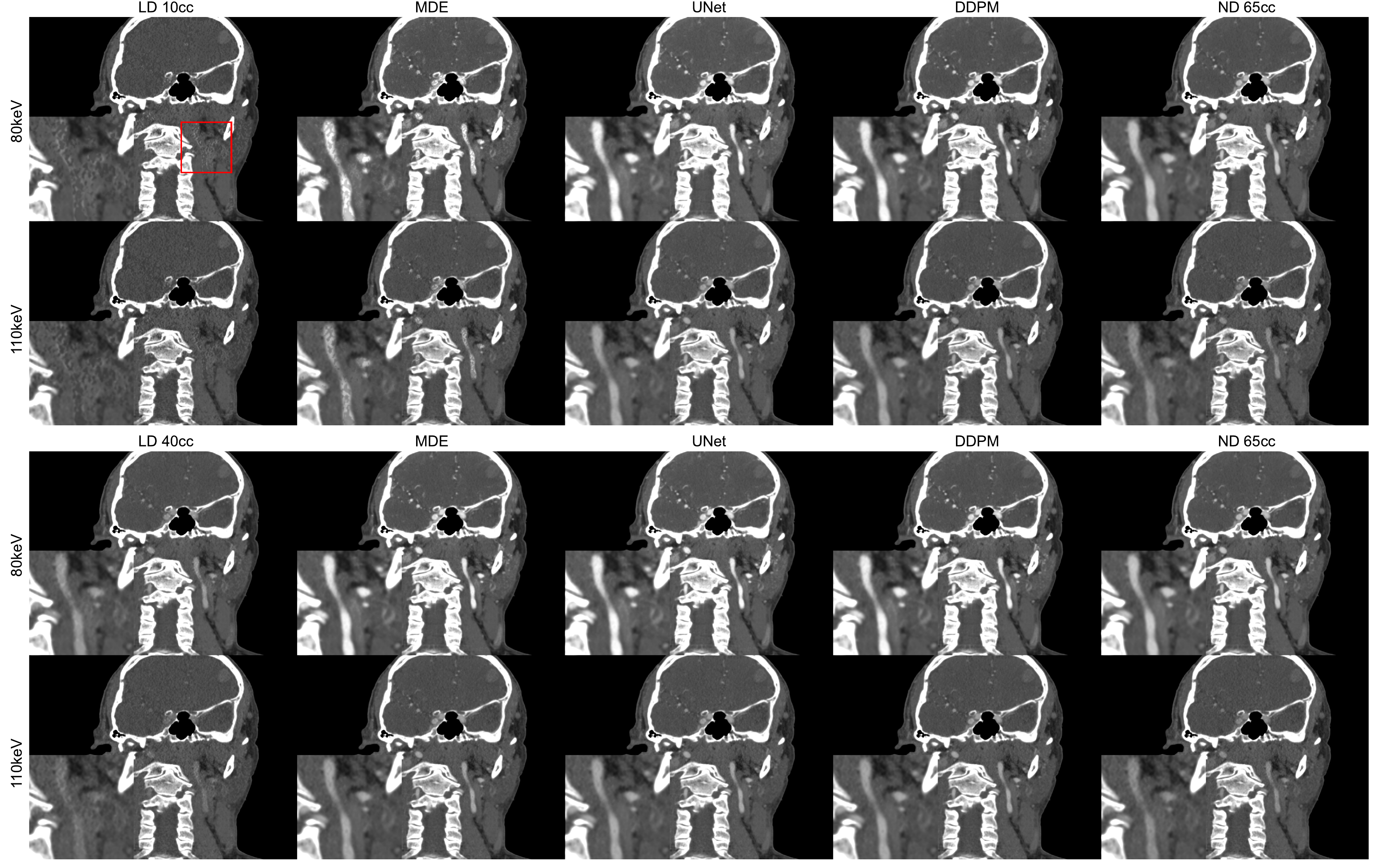}
\caption{Coronal view of contrast-enhancement results used for the reader study. Top: from left to right, 10 $cc$ simulated low contrast dosage images, MDE, UNet, DDPM contrast-enhanced results, and 65 cc normal contrast dosage images. Bottom: from left to right, 40 $cc$ simulated low contrast dosage images, MDE, UNet, DDPM contrast-enhanced results, and 65 $cc$ normal contrast dosage images. Bottom left corner of each sub-figure shows zoomed in region bounded by the red box.}
\label{fig_5}
\end{figure*}

\begin{figure*}[!ht]
\centering
\includegraphics[width=\textwidth]{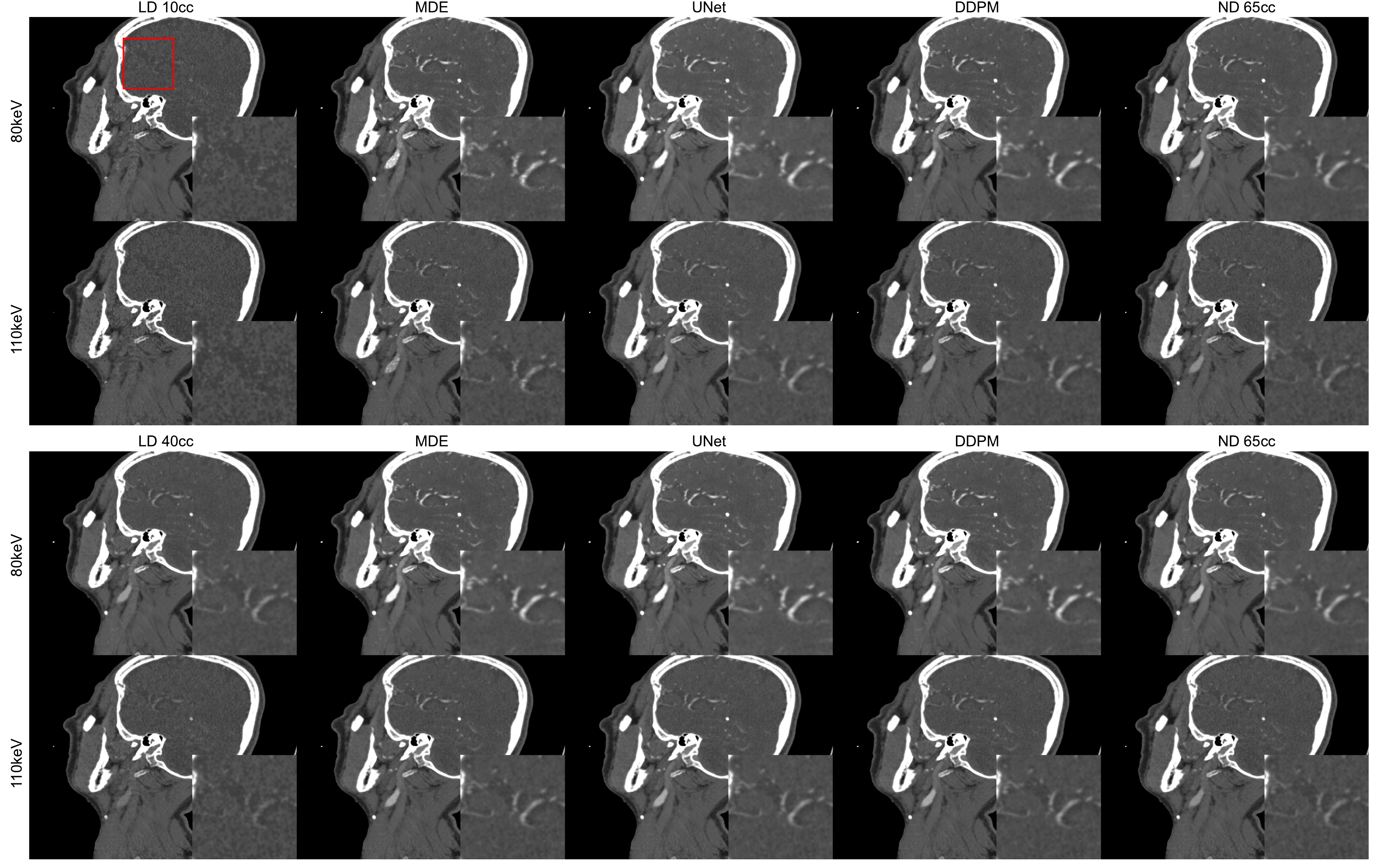}
\caption{Sagittal view of contrast-enhancement results used for the reader study. Top: from left to right, 10 $cc$ simulated low contrast dosage images, MDE, UNet, DDPM contrast-enhanced results, and 65 cc normal contrast dosage images. Bottom: from left to right, 40 $cc$ simulated low contrast dosage images, MDE, UNet, DDPM contrast-enhanced results, and 65 $cc$ normal contrast dosage images. Bottom left corner of each sub-figure shows zoomed in region bounded by the red box.}
\label{fig_6}
\end{figure*}

Figs.~\ref{fig_5} and~\ref{fig_6} demonstrate the coronal and sagittal views of a case used for the reader study. It can be found that blood vessels are barely shown in low contrast dosage images, especially for the 10 $cc$ setting. Meanwhile, all the three contrast enhancement methods can successfully demonstrate vessels in good contrast compared with corresponding low contrast dosage images. DDPM and UNet results overcome corresponding MDE results with smoother vessel blood pixel intensity demonstration.

\subsection{Reader Study Results}
We calculated the average score of normal cases, abnormal cases and all cases from each reader, as shown in Table~\ref{tab_2}. It can be observed that under the 10 $cc$ setting, the MDE method showed minimal improvement in image quality and achieved similar low scores as the low contrast dose images. On the other hand, the other two deep learning methods demonstrated significant improvements in image quality and produced favorable results. Under the 40 $cc$ setting, all the three contrast enhancement methods yielded substantially better results than the corresponding low contrast dose images. Notably, with our conditional DDPM method, we achieved contrast-enhanced results that exhibit similar scores to the normal dosage images.

\begin{table}[!t]
\caption{Reader study results using various methods for 10cc and 40cc contrast enhancement. \label{tab_2}}%
\centering
\begin{tabular}{c c | c c c c | c c c c | c}
\toprule
 & & \multicolumn{4}{c}{10 $cc$} & \multicolumn{4}{c}{40 $cc$} & \\
 & & LD & MDE & UNet & DDPM & LD & MDE & UNet & DDPM & ND \\
\midrule
\multirow{3}{*}{Reader 1} & Abnormal & 1.0 & 1.2 & 2.4 & 3.4 & 1.8 & 3.0 & 3.8 & 3.8 & 3.8 \\
 & Normal & 1.0 & 1.0 & 2.8 & 3.6 & 1.2 & 2.8 & 3.6 & 3.6 & 4.0 \\
 & Overall & 1.0 & 1.1 & 2.6 & 3.5 & 1.5 & 2.9 & 3.7 & 3.7 & 3.9 \\
 \hline \\
\multirow{3}{*}{Reader 2} & Abnormal & 1.0 & 1.6 & 2.4 & 2.8  & 1.6 & 2.6 & 2.4 & 2.8 & 3.2 \\
 & Normal & 1.0 & 1.2 & 2.8 & 3.4 & 1.6 & 3.0 & 3.4 & 3.6 & 4.0 \\
 & Overall & 1.0 & 1.4 & 2.6 & 3.1 & 1.6 & 2.8 & 2.9 & 3.2 & 3.6 \\ 
 \hline \\
\multirow{3}{*}{Reader 3} & Abnormal & 1.0 & 1.6 & 3.2 & 3.2 & 2.2 & 3.4 & 4.0 & 3.8 & 4.2 \\
 & Normal & 1.0 & 1.0 & 3.2 & 3.2 & 1.6 & 3.0 & 3.4 & 3.8 & 4.0 \\
 & Overall & 1.0 & 1.3 & 3.2 & 3.2 & 1.9 & 3.2 & 3.7 & 3.8 & 4.1 \\ 
\bottomrule
\end{tabular}
\end{table}

\section{Discussions}
Iodinated contrast media is extensively used in contrast-enhanced DECT scans to visualize vasculature. When the dosage of contrast media is reduced, vessels become vague and even disappear from DECT images. As a result, contrast-reduced images have lower quality, making it difficult to directly use them for uncompromised diagnostic performance. Table~\ref{tab_2} demonstrates a large gap between reader scores of low contrast dose images and normal contrast dose images. This gap becomes larger when the dosage is further reduced from 40 $cc$ to 10 $cc$. Using our proposed method, contrast of vessels can be enhanced and clearly presented, approaching a decent image quality similar to that of normal contrast dose scans, as shown in Figs.~\ref{fig_3} and ~\ref{fig_4}. Impressively, our approach can achieve good outcomes even in an extremely low contrast dose setting, even with only 12.5\% of the normal contrast dose. The reader study results in Table~\ref{tab_2} show that our results have similar scores to that for normal dose images.

When compared with the MDE and UNet methods, our method produces superior results that are close to the ground truth, as evidenced by excellent reader scores. Due to the interference of noise within DECT images and the oversimplification of the material decomposition hypothesis, the iodinated contrast component obtained using the three-material decomposition algorithm may not be most accurate. This inaccuracy is more serious in a lower contrast dosage setting. As a result, there are some pixels in the vessel region that are not decomposed as contrast, and the material decomposition result of iodinated contrast component contains some small holes in the vessel region using the MDE method.Using the UNet method, although the holes can be somewhat mitigated, they are still noticeable in the final results. In contrast, our method completely eliminates the holes and ensures that the results are very close to the normal contrast dose scans, as presented in Fig.~\ref{fig_4}. Since our method utilizes the same neural network architecture as UNet, the superior results must have come from the strong generative ability of the DDPM framework.

Our proposed method may significantly contribute to clinical tasks enabled by contrast-enhanced DECT scans from both healthcare and economic perspectives. 
Reducing contrast dosage means a reduced occurrence rate of iodinated contrast-induced adverse effects, such as acute kidney injury. We will conduct clinical studies to optimize our method for clinical translaton. Additionally, using lower contrast dose DECT can greatly reduce the consumption of contrast media, addressing the issue of contrast media 
shortage. Also, the potential economic contribution of our study is obvious. Taking the Atrium Health Wake Forest Baptist hospital as an example. The average cost of a typical Ominipaque 350 contrast media is around \$0.15 per milliliter in Winston-Salem, and it is estimated that around 3,200 liters of contrast media are used every year. It can be found that the annual contrast media cost is around \$480,000. By adopting the technique developed in this study, the amount of contrast media usage can be greatly reduces. Let's say that the reduction is just by half, the financial gain is already substantial, which around \$240,000 saved every year.

Our current study is based on simulation to obtain contrast reduced images. In the future, we will conduct studies on phantoms and animals such as non-human primates to obtain real contrast reduced DECT scan datasets. Then, we will conduct clinical trails and establish clinical feasibility and utilities of low contrast dose DECT and MRI in the future.

\section{Conclusion}
In conclusion, we have proposed a DDPM-based deep learning framework to relax the requirement of iodinated contrast media in DECT scans. We first use material decomposition to extract the iodinated contrast component and then create multi-level contrast dose reduced images. Conditional DDPMs are then trained to enhance the iodinated contrast component and generate contrast-enhanced images. According to our results, we can generate contrast-enhanced images with  image quality comparable to that of normal contrast dose images, even with a contrast dose as low as 12.5\% of the normal contrast dose. Our approach outperforms selected competing methods quantitatively and visually in our reader study. Further work is underway to optimize the approach for clinical translation.

\section*{Acknowledgments}
This work was partially supported by US National Institutes of Health grants R01CA237267, R01EB032716 and R01EB031885.

\bibliographystyle{unsrt}  
\bibliography{references}

\end{document}